\documentclass[]{article}
\usepackage{amsmath,amssymb,cite}
\usepackage{graphicx}
\makeatletter
\@addtoreset{equation}{section}
\makeatother

\newcommand{\ba}{\begin{eqnarray}}
\newcommand{\ea}{\end{eqnarray}}
\def\half{\frac{1}{2}}
\def\ben{\begin{equation}}
\def\een{\end{equation}}
\def\bea{\begin{eqnarray}}
\def\eea{\end{eqnarray}}
\def\be{\begin{equation}}
\def\ee{\end{equation}}

\def \bB {{\bf B}}

\def\nn{\nonumber}
\def\p{\partial}

\begin{document}
\def\ben{\begin{equation}}
\def\een{\end{equation}}
\def\bea{\begin{eqnarray}}
\def\eea{\end{eqnarray}}
\def \nn {\nonumber}
\def \bE {{\bf E}}\def \bB {{\bf B}}
\def \bx {{\bf x}}
\def \p {\partial}
\def \by {{\bf y}}
\def\half{{1\over 2}}
\def\ft#1#2{{\textstyle{\frac{\scriptstyle #1}{\scriptstyle #2} } }}
\def\fft#1#2{{\frac{#1}{#2}}}
\def\cA{{\cal A}}
\def\bm{\bibitem}
\def \bpi{\boldsymbol \pi}
\def\del{{\partial}}
\def\im{{{\rm i}}}

\def\Ared{{{\cal A}_{\rm red}}}

\newcommand{\hoch}[1]{$\, ^{#1}$}

\newcommand{\auth}{{\Large\bf{M. Cveti\v c\hoch{*,**}, 
G.W. Gibbons\hoch{\dagger}
and C.N. Pope\hoch{\ddagger,\dagger}}}}

\thispagestyle{empty}
\begin{flushright}
 \small{UPR-1284-T \ \ MI-TH-1751 \\}
\end{flushright}
\vspace{0.2in}

\begin{center}

{\Large{\bf STU Black Holes and SgrA$^\star$ }}

\vspace{22pt}
\auth

\vspace{30pt}{\hoch{*}\it Department of Physics and Astronomy,\\
University of Pennsylvania, Philadelphia, PA 19104, USA}

\vspace{10pt}{\hoch{\dagger}\it DAMTP, Centre for Mathematical Sciences,\\
 Cambridge University, Wilberforce Road, Cambridge CB3 OWA, UK}

\vspace{10pt}{\hoch{\ddagger}\it Mitchell Institute for Fundamental Physics
and Astronomy,\\
 Texas A\&M University, College Station,  TX 77843-4242, USA}

\vspace{10pt}{\hoch{**}\it Center for Applied Mathematics and Theoretical Physics,\\ University of
Maribor, SI2000 Maribor, Slovenia
}

\vspace {30pt} 
%\begin{document}

%
%\begin{abstract}

\underline{ABSTRACT}
\end{center}

The equations of null geodesics in the STU family of rotating black hole
solutions of supergravity theory, which may be considered as deformations of
the vacuum Kerr metric,  are completely integrable.
We propose that they be used as a foil to test, for example, 
with what precision the gravitational field  external to 
 the black hole at the centre of 
our galaxy is given by the  Kerr  metric. 
By contrast with some metrics  proposed  in the literature, the STU  metrics
satisfy by construction the dominant and strong energy conditions.  
Our considerations may be 
extended to include the effects of a  cosmological term. We show that these 
metrics permit a straightforward calculation of the
properties of black hole shadows.

%\end{abstract}
\newpage

\tableofcontents

\eject
%\end{center}

\section{Introduction}

Current and forthcoming observations of
the galactic centre, and in particular its shadow \cite{1,2,3},
should ultimately allow  a rather precise
determination of the metric around 
the black hole associated with the source ${\bf SgrA^\star}$.
Since the black hole is likely to be rotating 
at a near maximal rate, i.e. with the ratio
of angular momentum $|J|$ to mass $M^2$ 
close to unity, an important  goal of these observations is 
to check how accurately the metric agrees  
with that given by the Kerr solution \cite{Kerr:1963ud}  of the Einstein
equations. If this goal can be achieved one should have a 
powerful  observational test of the Einstein equations at 
the fully non-linear level (see, e.g., 
\cite{Johannsen:2015mdd,Johannsen:2015hib,Grenzebach:2014fha}). Moreover,
according
to a well known series of so-called No-Hair  theorems following 
rigorously from Einstein's equations the possible stationary metric 
should depend only on the two parameters $J$ and $M$.    
In making such a test it is thus  
desirable to have at hand  a family of metrics 
depending on say $n$ additional dimensionless  parameters $\lambda_i$,   
$i=1,2\dots ,n$,
such that if $\lambda_i=0$, we obtain the  Kerr solution.
The observational goal them becomes to determine
upper bounds on the magnitudes of the parameters $Q_i$. 
Of course this programme is not limited to the case
of ${\bf SgrA^\star}$ and applies to any 
plausible astrophysical black hole candidate but
its seems that it is ${\bf SgrA^\star}$ that offers the greatest 
promise of  progress in the near future \cite{Johannsen:2015mdd}.

Since  the observational signals  from black holes
depend to a large 
extent on  the behaviour of timelike and null geodesics
in the vicinity of the horizon, it is computationally highly desirable
that the geodesic equations for the family of metrics       
should possess the non-trivial property that
they  be integrable, as they are
in the Kerr case \cite{Carter:1968rr}.
Possible families of metrics for this purpose were  introduced
in \cite{Vigeland:2011ji,Johannsen:2011dh}, but they suffer from 
various disadvantages  and a 
more convenient family of metrics  was introduced 
in  \cite{Johannsen:2015pca}.  
These  have the merit of being expressible in a ``Kerr-Schild-like''  form,
which facilitates fully relativistic magnetohydrodynamic calculations.

Imposing the requirement that the Hamilton-Jacobi equation
\ben
 -\frac{\p S}{\p\tau} =   \ft12 g^{\mu \nu} \p _\mu S \p_ \nu S  \label{HJeqn}
\een
should have separable solutions  of the form
\ben
S= \half \mu ^2 \tau -Et  + L \phi +  S_r(r) + S_\theta (\theta)  
\label{massiveS}
\een
leads  to a family of metrics with four arbitrary
functions of $r$ and four arbitrary functions of $\theta$.
Imposing various regularity and asymptotic flatness 
conditions reduces this to just four functions of $r$
that have large-distance expansions in inverse powers of $r$. 

   This provides a powerful general framework
but it is perhaps too general, in the sense that
with such infinite-dimensional arbitrariness it may be possible
to fit almost any observational data. It  may therefore be 
necessary  to  restrict the range  of possibilities.
One such restriction would  arise from imposing
the dominant  and the strong energy conditions, so that the family of
metrics would at least be generated by physically-allowable matter
sources. However
this will still not substantially reduce the infinite-dimensional
manifold of possible metrics.

In this note we would like to suggest that
an interesting family  of  possibilities, not
completely overlapping with those 
introduced in  \cite{Johannsen:2015pca},
is contained within the class of much studied and completely explicit
asymptotically-flat
rotating black hole solutions  of 
STU supergravity theory \cite{Cvetic:1996kv,Chong:2004na}.  
These include the Kerr and Kerr-Newman 
metrics as special cases, and in general depend
on six parameters, two of which correspond
to the total energy $M$ and angular momentum $J$, with the 
other four associated to generalised charges $Q_i$ carried by gauge fields in
the supergravity theory. The most general  STU black holes solutions, with 
the additional (fifth) independent generalised charge parameter, 
were recently obtained  by Chow and Comp\`ere \cite{Chow:2013tia,Chow:2014cca}.

Some of these  possibilities within
the asymptotically-flat STU black hole solutions   have been considered in this
context already, including the Kerr-Newman metric, in \cite{Vries}, and   
the Kaluza-Klein rotating dilaton black hole, in
  \cite{Amarilla:2013sj,Amarilla:2015pga,Aliev:2013jya}.
Another  case which has been extensively investigated is
is the so-called Kerr-Sen black hole
\cite{Blaga:2001wt,Hioki:2008zw,Houri:2010fr}.
Because in this case the timelike geodesics
are also integrable, it has  also been the subject of studies
of the Banados-Silk-West (BSW) \cite{Banados:2009pr}   effect
\cite{Wei:2010gq,Debnath:2015bna}.

The integrabiity of the
timelike geodesics turns out to be a more general phenomenon.
In \cite{Chow:2013tia,Chow:2014cca} it was shown that
a wider class  of axisymmetric-stationary metrics admits separation for  
both
the null and timelike geodesics. Note, that this result is a consequence of 
the separability of the minimally coupled massless scalar equations in the  STU black hole background  \cite{Cvetic:1997xv,Chow:2013tia,Chow:2014cca}.  

The Chow-Comp\`ere  solutions  include the most
general seed from which to generate, using an $SO(4,4)$ solution
generating procedure, all rotating charged solutions of ${\cal N}=8$
un-gauged supergravity theory provided one works in string conformal frame
rather than Einstein-conformal frame.
The two conformal frames in the case of the Sen Black holes
In fact comparing equation (5.15) of  \cite{Chow:2014cca}
with equation (17) of \cite{Johannsen:2015pca} one finds
agreement in the case of null geodesics. See also 
\cite{Aliev:2013jya} in the Kaluza-Klein case.
This observation of Chow and Comp\`ere opens the way
to a much more extensive comparison of astronomical observations
and discussions of such topics as the BSW effect with the STU metrics.

Some other cases not directly related to STU supergravity include
the rotating braneworld black hole 
\cite{Amarilla:2011fx}, and  the rotating black hole in
extended Chern-Simons modified gravity \cite{Amarilla:2010zq} 
for rotating black holes with exotic matter \cite{Tinchev:2015apf}. 
For a review
of some of these see   \cite{Amarilla:2015pgp}.

The paper is organised as follows.

In section 2 we introduce the 4-charge STU
family of charged rotating  black holes and describe their
separability properties.
They depend on 4 real charge parameters $\delta_i$, in addition to
the mass and rotation parameters.  We present  formulae
for the ADM mass, total angular momentum, four charges and
and four magnetic dipole moments. We show that these satisfy
some interesting inequalities. By specialising the parameters
 we obtain various well-known special cases, some of which have
already been studied in the astrophysical literature.  In the non-rotating case
we give the values of the Parametrised Post Newtonian (PPN) parameters
$\beta$ and $\gamma$ for spherically-symmetric metrics in terms
of the parameters $\delta_i$. 

In section 3  we discuss null geodesics and in section 4 equatorial
geodesics, both timelike and null.

Section 5 is devoted to the study of shadows.
We give  a general discussion and then specialise to
the case where the four electric charges are set pairwise equal, for which
the problem can be reduced 
to a straightforward numerical procedure.  We have collected some details
about the pairwise-equal charged black holes of STU supergravity in an
appendix. 

\section{Rotating, Asymptotically  Flat STU Black Holes} 

\subsection{The STU metrics}

In this section we shall focus on rotating black holes in ungauged STU 
supergravity, characterised by the  mass the  $M$, the angular momentum 
$J$ and four charges $Q_i$.  Employing solution-generating techniques, 
these metrics were first obtained in
\cite{Cvetic:1996kv}, 
with all the sources  explicitly displayed  in  \cite{Chong:2004na}.
Here we shall present  the metric only:\footnote{For the  STU 
Lagrangian of  ${\cal N}=2$ supergravity coupled to three vector 
supermultiplets, and the  explicit form of the full solution, 
see \cite{Chong:2004na}.  
See also, \cite{Cvetic:2011,Cvetic:2012,Cvetic:2014eka}.
The black hole solutions of the STU theory are generating 
solutions of ${\cal N}=4$ and ${\cal N}=8$  supergravity theory, which 
can be obtained as a  toroidal compactification on an effective 
heterotic string theory and Type IIA superstring theory, respectively. The 
full set of solutions of these maximally supersymmetric supergravity theories 
can be obtained by acting with a subset of  
respective $\{S,T\}$-  and $U$- duality transformations. 
(See e.g., \cite{CT, CH}.)} 
%%%%%
\bea
d{ s}^2_4 & =& -{\Delta}^{-1/2}\, 
   { G} ( d{ t}+{ {\cal  A}})^2 + { \Delta}^{1/2}
\Big( \fft{dr^2}{X}  + d\theta^2 + {{ X}\over{  G}} 
\sin^2\theta d\phi^2\Big)~,\label{metricg4d}
\eea
where 
\bea
{ X} & =& { r}^2 - 2{ m}{ r} + { a}^2~,\cr
{ G} & = &{ r}^2 - 2{ m}{ r} + { a}^2 \cos^2\theta ~, \cr
{ {\cal A}} & =& {2{ m} { a} \sin^2\theta \over { G}}
\left[ ({ \Pi_c} - { \Pi_s}){  r} + 2{ m}{ \Pi}_s\right] d\phi~,
\eea
and
\bea
{\Delta} =&& \prod_{i=1}^4 ({ r} + 2{ m} s_i^2) +a^4 \cos^4\theta \nn\\
&&+ 2 { a}^2 \cos^2\theta [{ r}^2 + { m}{ r}\sum_{i=1}^4 s_i^2
+\,  4{ m}^2 ({ \Pi}_c - { \Pi}_s){ \Pi}_s   -  2{ m}^2 \sum_{i<j<k}
 s_i^2 s_j^2 s_k^2 ] \,.  \label{Delta} 
\eea
We are employing the following  abbreviations:
\ben
{ \Pi}_c \equiv \prod_{i=1}^4 c_i\,,\qquad
\ { \Pi}_s \equiv  \prod_{i=1}^4 s_i\,,\qquad
 s_i=\sinh\delta_i\,,\qquad c_i=\cosh\delta_i~.\label{pics}
\een
The solution is parametrised by the bare mass parameter $m$, the rotational 
parameter $a$ and four charge parameters $\delta_i$ $(i=1,2,3,4)$.  The 
solution is written as a  fibration over a three-dimensional base
that is itself independent of the charge parameters, and with a warp factor 
denoted by  ${\Delta}$.

   Defining 
%%%%% 
\be
\Ared = 2m [(\Pi_c-\Pi_s) r +2m \Pi_s]\,,\qquad
\nu= \fft{\Delta - \Ared^2}{G}\,,\label{Arednudef}
\ee
%%%%%
and noting that $G=X - a^2 \sin^2\theta$, it can be seen that the inverse
of the metric (\ref{metricg4d}) is given by \cite{Cvetic:2011}
%%%%%
\bea
\Delta^{1/2}\, \Big(\fft{\del}{\del s}\Big)^2 =
  -\fft1{X}\, \Big( \Ared\, \fft{\del}{\del t} + a\fft{\del}{\del\phi}
\Big)^2
   + X\, \Big(\fft{\del}{\del r}\Big)^2 + \Big(\fft{\del}{\del\theta}\Big)^2
 + \fft1{\sin^2\theta}\, \Big(\fft{\del}{\del\phi}\Big)^2
   - \nu\, \Big(\fft{\del}{\del t}\Big)^2\,,\label{inversemet}
\eea
%%%%%

\subsection{Electric charges, angular momentum and magnetic dipole moments}

The electromagnetic properties of 
rotating electrically charged STU black holes are reviewed in 
\cite{Cvetic:2013roa}.  Employing the notation introduced above the mass $M$, 
angular momentum $J$, the four electric charges $Q_i$ and the four 
induced dipole charges $\mu_i$  are given by\footnote{Here we are using
the normalisation of the four $U(1)$ gauge fields in which, in the
absence of scalar fields, the Lagranging has the form ${\cal L} \sim
\sqrt{-g} (R-\ft14 F_i^2)$, with the charges being of the form
$Q_i\sim 1/(4\pi) \int {*F_i}$.  This is to be contrasted with what we
refer to later as the ``canonical normalisation'' for an electromagnetic
field, for which the Lagrangian has the form ${\cal L}\sim \sqrt{-g} (R-F^2)$
and the charge is defined as $Q=1/(4\pi)\int {*F}$.}
%These depend upon the 4 boost parameters $\delta_i$.
%If $c_i= \cosh \delta_i$, $s_i= \sinh \delta_i$,  $\Pi_c=c_1c_2c_3c_4$,
%$\Pi_s=s_1s_2s_3s_4$, $\Pi_c^1= c_2c_3c_4\, {\rm etc}$,
%$\Pi_s^1= s_2s_3s_4\, {\rm etc}$,  then according to \cite{Cvetic:2013roa}
\bea
\qquad M= \frac{m}{4} \sum _{i=1}^4 \cosh 2\delta_i  \,,\qquad
J&=& ma \big( \Pi_c -\Pi_s \bigr)\\ 
Q_i = m \sinh 2\delta_i \,,\qquad 
\mu_i  &=&  2ma \bigl(s_i \Pi^i_c - c_i  \Pi^i_s
\bigr ) \,, \label{CY}
\eea
where $c_i$, $s_i$, $\Pi_s$ and $\Pi_c$ are defined in (\ref{pics}), 
$\Pi_c^i= \Pi_c/c_i$ and 
$\Pi_s^i= \Pi_s/s_i$. 
Evidently
\ben
4M \ge \sum_i |Q_i| \,. \label{BOG}
\een

One also has
\ben
\frac{\mu_i}{2J}  = \frac{s_i \Pi ^i_c- c_i \Pi^i_s}{\Pi_c-\Pi_s}\,. 
\een
If we assume that $s_i>0$ for all $i$, then 
\ben
\frac{\mu_i}{2J} \le \frac{ s_i(\Pi ^i_c- \Pi^i_s)}{c_i (\Pi^i_c-\Pi^i _s)} 
\le \tanh \delta _i \le 1\,. \label{BOUND}
\een

There are some special cases which coincide with
solutions of an Einstein-Maxwell-Dilaton theory for particular choices
of the dilaton coupling to the electromagnetic field:
%%%%%
\bigskip

\noindent{\bf Einstein-Maxwell Black Holes}:
\medskip

Here $\delta_1=\delta_2=\delta_3=\delta_4=\delta$.
Thus from (\ref{CY}) 
\bea
M= m \cosh{2 \delta} \,,\qquad J&=& ma \cosh 2 \delta \\ 
Q_i= m \sinh 2 \delta  \,,\quad \mu_i  &=& ma  \sinh  2 \delta
\eea
and hence if we set $Q=Q_i$  and $\mu=\mu_i$, corresponding to a
canonically-normalised electromagnetic field $F=F_i$ such that the 
Lagrangian is ${\cal L}= \sqrt{-g}(R-F^2)$, we find that $g=2$ and  
\ben
\frac{|\mu|}{|J|}  = \tanh  2 \delta = \frac{|Q|}{M} \le 1 \,.
\een
%%%%%
\bigskip

\newpage

\noindent{\bf Kerr-Kaluza-Klein Charged Black Holes}:
\medskip

Here $\delta_1=\delta$, $\delta_2=\delta _3=\delta_4 =0$, and so
%%%%%
\bea
M= \frac{m}{4} (3 +   \cosh{2 \delta})  ,\qquad J&=& ma \cosh  \delta \\
Q_1= m \sinh 2 \delta  \,,\qquad \mu_1 &=&  2ma \sinh \delta \,.
\eea
%%%%%
If we set $Q=\ft12 Q_1 $     and $\mu = \ft12 \mu_1$, corresponding
to a canonically-normalised electromagnetic field $F =\ft12 F_1$, we find that 
\ben
\beta = \frac{|\mu|}{|J|} =  \tanh\delta \le 1\,.  
\een

To obtain agreement with \cite{Gibbons:1985ac,Frolov:1987rj}
 we set 
\ben
v=\tanh \delta \,.
\een
We also have
\ben
M \ge \half |Q| 
\een
%%%%%
which is consistent with the general Bogomolnyi inequality
for Einstein-Maxwell Dilaton black holes \cite{Gibbons:1993xt}. 
\bigskip

\noindent{\bf Kerr-Sen Black Holes \cite{Sen:1992ua}}:
\medskip

 Here $\delta_1=\delta _3 =\delta$, $\delta_2=\delta_4=0$.  Thus
%%%%% 
\bea
 M = \half m (1+\cosh 2 \delta) \,,\qquad J&=& \half ma (1+ \cosh 2 \delta)
  \\
 Q_1=Q_2 = m \sinh 2 \delta \,,\qquad \mu_1=\mu _2
  &=&  ma \sinh 2\delta\,. 
 \eea
%%%%%
If we set $Q= (\sqrt{2})^{-1}\,  Q_1= (\sqrt{2})^{-1}\,  Q_3 $
and $\mu = (\sqrt{2})^{-1}\, \mu_1= (\sqrt{2})^{-1}\,  \mu_3 $, 
corresponding to a canonically-normalised electromagnetic field
$F=(\sqrt 2)^{-1}\, F_1=(\sqrt2)^{-1}\, F_3$, then
%%%%%
\ben
 \frac{|\mu|}{|J|} = \sqrt{2} \tanh \delta \,,
 \een
and  
 \ben
\frac{|Q|}{M} = \sqrt{2} \tanh\delta \le \sqrt{2} \,.  
\een
We find that that the gyromagnetic ratio is $g=2$, and we find
consistency with the general Bogomolnyi inequality
for Einstein-Maxwell Dilaton black holes \cite{Gibbons:1993xt}
and  the Kerr-Sen black hole \cite{Sen:1992ua}, provided  one  replaces
Sen's  $\alpha$  by $2 \delta$. 
\bigskip

\noindent{\bf Black Holes with Pairwise-Equal Charges}:
\medskip

   A further subclass class of solutions which affords considerable
simplifications arises when the  four charges
are taken to be pairwise equal, with
$\delta_3=\delta_1$ and $\delta _4= \delta_2$.  We give an 
explicit expression for the Lagrangian describing this truncation, and
we discuss also its duality symmetries and give the black hole solutions,
in appendix A.  After making the rescalings so that the two
$U(1)$ gauge fields have the canonical normalisations, as given in the
appendix, we see that
%%%%%
\be
J= a M =  \fft{m a}{2}\, (\cosh 2\delta_1 + \cosh 2\delta_2)\,,\qquad
\mu_i = a Q_i = \fft{m a}{\sqrt2}\, \sinh 2\delta_i\,,
\ee
%%%%%
and so
%%%%%
\be
\fft{|\mu_i|}{|J|} = \fft{|Q_i|}{M} = 
\fft{\sqrt2\, \sinh2\delta_i}{\cosh2\delta_1+\cosh2\delta_2} \le 
  \sqrt2\, \tanh\delta_i\,.
\ee
%%%%%

  The pairwise-equal charge solutions encompass the Kerr-Sen solution
if we set $\delta_2=0$ (and hence $\delta_4=0$ also).  They also 
specialise to the Kerr-Newman case if we set $\delta_1=\delta_2$ (and hence 
equal also to $\delta_3$ and $\delta_4$).

\subsection{Parametrised Post-Newtonian parameters}

We  follow the analysis of \cite{Abishev:2015pqa}.
The PPN or Eddington parameters $\beta$ and $\gamma$ are defined for
asymptotically flat  metrics by 
\bea
-g_{00}&=&  1-2  \frac{M}{|{\bf x}|} + 2 \beta \frac{M^2}{|{\bf x}|^2} +
  {\cal O}(\frac{M^3}{|{\bf x}|^3})\\ 
  g_{ij} & =&\delta _{ij} \bigl( ( 1+ 2\gamma \frac{M}{|{\bf x}|}  +
    {\cal O}(\frac{M^2}{|{\bf x}|^2}) \bigr ) \,.  
\eea

We recall \cite{Cvetic:2014vsa} that the static STU metrics may be cast
in isotropic coordinates as
\ben
ds^2 =- \bigl (1- \frac{m^2}{4 \rho^2} \bigr )^2  \Pi ^{-\half} dt^2
+ \Pi ^\half \Bigl(d  \rho ^2 + \rho ^2
\bigl (d \theta ^2 + \sin ^2 \theta d \phi ^2 \bigr ) \Bigr )   
\een
where
\ben
\Pi=  
\prod_i \Bigl (1+ \frac{m \cosh 2 \delta_i}{\rho }  + \frac{m^2}{4 \rho^2} 
\Bigr ) \een
whence
\ben
\Pi \approx 1+ \frac{4M}{\rho} + \frac{m^2}{\rho^2 } \sum_{i<j}
\cosh 2 \delta_i \cosh 2 \delta _j + \frac{m^2}{\rho ^2} \,, 
\een
%%%%%
and so
%%%%%
\ben
-g_{00} = \bigl(1-\fft{m^2}{4\rho^2}\bigr)^2\, \Pi^{-\ft12} 
  = 1- \frac{2M}{\rho}  + \frac{6M^2}{\rho^2} - \frac{m^2} {\rho^2}
- \frac{m^2} {2 \rho^2} \sum_{i<j}
\cosh 2 \delta_i \cosh 2 \delta _j \,.
\een
Thus 
%%%%%
\ben
\gamma=1\,,\qquad M= \frac{m}{4} \sum_i \cosh 2 \delta_i\,,
\een
%%%%%
and
%%%%%
%%%%%
\ben
\beta = 3 - \frac{8}{\Bigl( \sum_i \cosh 2 \delta_i \Bigr )^2 } -
 \frac{ 4 \sum_{i<j}
  \cosh 2 \delta_i \cosh 2\delta_j}{\Bigl( \sum _k \cosh2  \delta_k \Bigr) ^2}
\,.\label{beta0}
\een
%%%%%
This can be rewritten as
%%%%%
\be
\beta = 1 + \fft{ 2 \sum_i \sinh^2 2\delta_i}{\Bigl(\sum_j 
                  \cosh 2\delta_j\Bigr)^2}\,.
\label{beta}
\ee
%%%%%

In comparing (\ref{beta}) with
observational/experimental constraints on $\beta$,  
one should bear in mind that
these are typically derived in the case the when the gravitating body
is a star or planet rather than a black hole.

\section{Hamilton-Jacobi Equation and Null Geodesics}

\subsection{Hamilton-Jacobi equation in the STU black hole background}

 Taking $S$ in the Hamilton-Jacobi equation (\ref{HJeqn}) to
have the form $S= \half \mu^2\, \tau  -Et + J \phi + W(r,\theta)$, one sees
from the inverse metric (\ref{inversemet}) for the STU 
black holes that
%%%%%
\ben
- \nu  E^2 -\fft1{X}\,  (E \Ared - a j)^2 + \fft{j^2}{\sin^2\theta} 
+ X (\p_r W)^2  + (\p_\theta W )^2   = - \Delta^{1/2}\, 
\mu^2 \,,  \label{HJ}
\een
%%%%%
 where $(\mu,E,j)$  are rest mass, conserved energy and angular momentum
of the particle respectively.
The quantities $X$ and $\Ared$ depend only on $r$, while
$\nu$, which is a function of $r$ and $\theta$, is in fact the sum of
a function or $r$ and a function of $\theta$:
%%%%%
\ben
\nu (\theta, r)  = {\nu _\theta } (\theta) + {\nu _r } (r) \,,
\een
%%%%%
with
%%%%%
\bea
\nu_r &=& r^2 + 2m \big(1+\sum_i s_i^2\big)\, r + 8m^2\, \big(\Pi_c-\Pi_s\big)
\, \Pi_s - 4 m^2\, \sum_{i<j<k} s_i^2 s_j^2 s_k^2\,,\nn\\
\nu_\theta &=& a^2 \, \cos^2\theta\,,\label{massiveHJ}
\eea
%%%%%
It therefore follows that, as was shown in \cite{Keeler:2012mq}, the
Hamilton-Jacobi equation for massless geodesics in the 
4-charge STU black hole background separates.\footnote{The separation
of the Hamilton-Jacobi equation for null geodesics is in fact a
corollary of the previously discovered
fact \cite{Cvetic:1997xv,Cvetic:2011}  that the massless
wave equation is separable for all metrics in the family.
It is interesting to
note that the separability of the wave equation for a massless
scalar field also holds for the most general five-dimensional
STU black holes  \cite{Cvetic:1996xz}, as shown in
\cite{Cvetic:1997uw} , as well as for those with a cosmological
constant \cite{Wu}, as shown in \cite{Birkandan:2014vga}.}  

   As can be seen from (\ref{Delta}), for a general 4-charge STU black hole
the function $\Delta^{1/2}$ 
appearing in (\ref{massiveHJ}) will be an irrational
function of $r$ and $\cos\theta$, and thus the Hamilton-Jacobi 
equation for massive geodesics will not be separable.  However, in the
case of the pairwise-equal charge black holes, given in appendix A, 
$\Delta$ becomes a perfect square and in fact one then has
%%%%%
\be
\Delta^{1/2} = (r+ 2m s_1^2)(r+2m s_2^2) + a^2\, \cos^2\theta\,.
\ee
%%%%%
It is then evident, since $\Delta^{1/2}$ is a sum of a function of $r$
and a function of $\theta$, that the Hamilton-Jacobi equation is now 
separable
also for massive geodesics.  Of course this will continue to be the case also
for the specialisations of the pairwise-equal charge black holes to Kerr-Sen 
or to Kerr-Newman black holes.

\subsection{Null geodesics}

   Setting $\mu=0$ in (\ref{HJ}), we see that the Hamilton-Jacobi 
equation for null geodesics separates if we write
%%%%%
\ben
W(r,\theta)= R(r) + \Theta(\theta)\,,
\een
%%%%%
leading to the ordinary differential equations 
%%%%%
\bea
X \bigl(\frac{dR}{dr} \bigr )^2   - \nu_r  E^2 - \frac{1}{X}
(E\, \Ared - a j)^2 &=&K
\\
\bigl( \frac{d \Theta}{d \theta}  \bigr ) ^2   
+\frac{j^2}{\sin ^2 \theta }      -\nu_\theta  E^2 &=&-K \,,
\eea
%%%%%
where the separation constant $K$ corresponds to the well-known
Carter constant in the case of the uncharged (Kerr) black hole.
The geodesic equation $\dot x^\mu  = g^{\mu\nu}\, \del_\nu S$ then
gives
%%%%%
\bea
\Delta\, \dot r^2 &=& (E \Ared- a j)^2 + 
                (\nu_r \, E^2 + K)\, X\,,\nn\\
\Delta\, \dot\theta^2 &=& a^2\,E^2\,  \cos^2\theta -K -\fft{j^2}{\sin^2\theta}
\,,\nn\\
\Delta^{1/2}\, \dot\phi &=& \fft{ a(E \Ared -a j)}{X}  + 
   \fft{j}{\sin^2\theta}\,,\nn\\
\Delta^{1/2}\, \dot t &=& (\nu_r + a^2\, \cos^2\theta)\, E +
   \fft{\Ared}{X}\, (E \Ared - a j)\,.\label{geodesics}
\eea
%%%%%

\section{Motion in the Equatorial Plane}

This can be treated analytically  for both massive as well
as massless particles
even if the general motion out of the plane is not integrable,
since it reduces to a one-dimensional problem. 
We begin by noting from the general metric (\ref{metricg4d}) that
the horizon, which is a null hypersurface,
is  located at the larger zero, $r=r_+$,  of $X$.   Furthermore the
ergosurface, which is a timelike hypersurface on which the
Killing vector $\frac{\p}{\p t}$ becomes null,  is given by $G=0$. 

One may consistently restrict the motion to the equatorial
plane $\theta =  \frac{\pi}{2} $, since this is totally geodesic by virtue
of being fixed under the reflection symmetry
$\theta \rightarrow \pi - \theta$. 

Taking into account the conservation of energy $E$ and angular momentum $j$,
it follows from (\ref{geodesics}) that the radial equation reduces to 
\bea
\Delta\, \Bigl( \frac{dr}{d \lambda} \Bigr )^2 &=&
X (\nu_r E^2 -\Delta ^\half \mu ^2  -j^2 ) + ( E \Ared  -a j )^2\\
&=& (\Ared^2 + \nu_r X) (E- V^+(r))(E-V^-(r)) \,.  
\eea
%%%%%
The ``effective potentials ``  $V^\pm (r)$ depend upon the mass 
parameter $m$, the charge
parameters $\delta_i$, the rotation parameter $a$  and the impact parameter 
$j$.

It is convenient to plot, at least in one's mind,
the two branches $E=V^\pm(r)$. The motion may be envisaged as a horizontal
line in the $E-r$ plane with turning points when the straight line intersects
one of the two branches $E=V^\pm(r)$. The two branches correspond
to prograde or retrograde  motion. At infinity $V^\pm$ limits to $\pm \mu$,
where $\mu$ is the mass of the particle, and they coalesce at the 
horizon $r=r_+$,  
at which
\ben
E=E_+= V^+(r_+)=V^-(r_+) = \frac{a J}{\Ared(r_+) }
\een

 Circular orbits correspond to local
maxima or minima of $V^\pm(r)$. For $V^+(r)$ the maxima are unstable
and the minima stable. The opposite is true for $V^-(r)$.

\section{Shadows and Superficial  Orbits}

It was recognised early on in the subject
\cite{Bardeen1,Bardeen2,Bardeen:1973tla} that
among the null geodesics were  a family that
stayed at constant values of the Boyer-Lindquist coordinate $r$.
These are sometimes   referred to as spherical geodesics
\cite{Teo,Grenzebach:2014fha}  
or   circular geodesics or even  closed geodesics \cite{Vries}. None of these   terms is entirely 
appropriate.  Unlike  the spherically symmetric non-rotating  case,
where not only is this  a single round 2-sphere of constant curvature
on which the null geodesics project to great circles, and so
are  necessarily
closed and of constant curvature,
in the rotating  case there is no single radius, but rather a range
of radii and     none of the  2-surfaces $r=\,$constant
onto which the null geodesics project are  of constant curvature.
Moreover the projected curves are not, unless special directions are chosen,
closed curves, and even then, not of constant curvature. Finally, while
it is obvious in the spherical case that every great  circle
may be regarded as the projection of a unique null geodesic, this
is not obviously true in the rotating case.

In  \cite{Gibbons:2016isj}  examples were given of static spacetimes
admitting  spacelike spherical 2-surfaces such that every geodesic lying
in the 2-surface may be regarded as the projection of a unique
null geodesic. In \cite{Cvetic:2016bxi} it was proposed
that such surfaces, which are totally geodesic submanifolds
of a suitably defined optical metric, be termed photon surfaces
or anti-photon surfaces, depending upon whether the null
geodesics were unstable or stable respectively. Because of the
misgivings we have voiced above, in what follows we shall refer to null
geodesics lying in a spacelike 2-surface as {\it superficial orbits},
and the surfaces as photon surfaces or anti-photon surfaces.  

The first discussion of the shadow phenomenon was by Synge
\cite{Synge}.  In effect, he pointed out that
a compact spherical  body surrounded by or enclosing
a photon sphere would cast a shadow on the night sky, subtended by
photons entering an observer's eye that had just grazed the photon sphere.
The discussion in the rotating case is similar, although
the shadow is no longer circular, nor central with
respect to the line of centres even if the observer is located
in the equatorial plane of the black hole. 
We commence by noting that the  general four charged black
rotating black holes of the STU theory share an important
simplifying feature that holds for photons crossing
the equator in  the Kerr case.  Our discussion will 
follow the lucid papers  of Bardeen and Teo \cite{Bardeen:1973tla,Teo}.

\subsection{Shadow theory}

The basic strategy for finding the shadow was described, for example, in
\cite{Bardeen:1973tla}.  We shall follow a more recent discussion in
\cite{Vazquez:2003zm}.   An observer at a large distance
$r=r_0$ from the black hole, at co-latitude $\theta_0$, can set up 
a Cartesian coordinate system $(\alpha,\beta)$ in the plane orthogonal
to the line from the black hole to the observer.  Thus every photon 
reaching the  observer defines a point in the 
$(\alpha,\beta)$ plane.
The totality of such points fill out a region representing  the 
observer's unobstructed night sky.
Its complement is the the shadow of the black hole, and we seek its boundary.
This consists of photon orbits that just fail to escape
falling into the black hole.   These are the solutions of
%%%%%
\ben
R(r)=0\,, \qquad R^\prime (r) =0 \,,\label{orbitcon}
\een
%%%%%
subject to the requirement that the orbits are unstable. 
For each such superficial orbit, one can calculate the corresponding 
point in the $(\alpha,\beta)$ plane, as described below:

In general the values of $\alpha$ and $\beta$ for a light ray
reaching the observer are related to
the asymptotic values of $d\phi/dr$ and $d\theta/dr$ by  \cite{Vazquez:2003zm}
%%%%%
\be
\alpha= -r_0^2\, \sin\theta_0\, \fft{d\phi}{dr}\Big|_{r=r_0}\,,\qquad
\beta = r_0^2\, \fft{d\theta}{dr}\Big|_{r=r_0}\,.
\ee
%%%%%
 From (\ref{geodesics}), and from the asymptotic forms of the various
metric functions, we therefore find
%%%%%
\be
\alpha =-\fft{\lambda}{\sin\theta_0}\,,\qquad
\beta = \big[ \eta + a^2\, \cos^2\theta_0 - 
   \lambda^2\, \cot^2\theta_0\big]^{1/2}\,,\label{alphabeta}
\ee
%%%%%  
where we have defined the {\it impact parameters} 
%%%%%
\ben
\lambda = j/E \,, \qquad \eta = Q/E^2 \,.
\een
%%%%%
Here, $Q$ is defined such that the additive constant 
in the $\dot\theta^2$ equation in (\ref{geodesics}) 
vanishes on the equatorial plane $\theta=\frac{\pi}{2}$,
i.e. 
%%%%%
\be
Q=-K-j^2\,.\label{Qdef}
\ee
%%%%%

   Imposing the conditions (\ref{orbitcon}) for superficial orbits around
the black hole allows one to solve for the impact parameters 
$(\lambda, \eta)$ as functions of the orbital radius $r$, where the 
allowed range of $r$
runs between the values $r_{\rm inner}$ and $r_{\rm outer}$ of the 
two circular
geodesics that lie in the equatorial plane. Using this information, 
and the relations (\ref{alphabeta}), one can plot
a curve in the $(\alpha,\beta)$-plane.  This curve defines the boundary of
the shadow cast by the black hole.

\subsection{Orbits crossing the equator with $\dot\phi=0$}

   Teo \cite{Teo}  observes that if he considers the spherical orbit 
at the radius that 
maximises his Carter constant $Q$, it has the property that it has $\dot\phi=0$
at the moment when it crosses the equator.  We find that this generalises
to all the STU black holes (including all Chow-Compere examples).  Here,
we show this for the 4-charge black holes discussed above.

   We can scale the energy $E$ of the particle by scaling the affine parameter.
We shall assume now that we scale it so that $E=1$, without loss of
generality.  By analogy with Teo, we define $Q$ to be the quantity
such that in the $\theta$ equation we have
%%%%%
\be
\Delta\, \dot\theta^2 = Q + {\cal O}(\cos^2\theta)\,,
\ee
%%%%%
i.e. $Q$ is the value of $\Delta\, \dot\theta^2$ on the equator.  Thus we see
from (\ref{geodesics}) that
%%%%%
\be
   Q= -K -j^2\,.
\ee
%%%%%

The radial equation is then $\Delta\, \dot r^2= H(r)$ with 
%%%%%
\be
H(r)= (\Ared - a j)^2 + (\nu_r - Q -j^2)\, X\,.
\ee
%%%%%
The conditions for spherical orbits, $H=0$ and $H'=0$, can be viewed as
determining $j=j(r_0)$ and $Q=Q(r_0)$ in terms of the orbital radius $r_0$. 
Thus we shall have
%%%%%
\bea
H &=& (\Ared -a j)^2 + (\nu_r -Q -j^2)\, X =0\,,\nn\\
H' &=& 2(\Ared -a j)\, \Ared' + \nu_r' \, X +
    (\nu_r -Q -j^2)\, X'=0\,,\label{HHp}
\eea
%%%%%
where all quantities are evaluated at the orbital radius $r_0$.

   We now look for the value of $r_0$ that maximises $Q$.  Thus we may now
differentiate the $H=0$ equation in (\ref{HHp}) 
 with respect to $r_0$, where now
we include the differentiation of $j(r_0)$ and $Q(r_0)$.  Since we then
require $Q'(r_0)=0$ at its maximum, this means we get an equation like the 
$H'=0$ equation in (\ref{HHp}), except with $j'$ terms also. Thus:
%%%%%
\be
2(\Ared -a j)\, \Ared' -2a (\Ared - a j)\, j' 
  + \nu_r' \, X  - 2j j'\, X 
+(\nu_r -Q -j^2)\, X' =0\,.
\ee
%%%%%
Subtracting the $H'=0$ equation in (\ref{HHp}), and assuming $j'(r_0)\ne 0$,
we therefore find
%%%%%
\be
a\, (\Ared -a j)  + j X=0\,,\label{zerophidot}
\ee
%%%%%
for the case of the spherical orbit whose radius maximises $Q(r_0)$.  It is
easy to see from the $\dot\phi$ equation in (\ref{geodesics}) that 
eqn (\ref{zerophidot}) implies that $\dot\phi=0$ when $\theta=\pi/2$.  In 
other words, the spherical orbits at the radius that maximises $Q$ 
always have the property that they cross the equator with zero velocity in the
$\phi$ direction. 

\subsection{Pairwise-equal charges}

   Considerable simplifications occur if the charges in the four-charge
rotating black hole solutions are set pairwise equal.  These
solutions are presented explicitly in (\ref{pairwisesol}). 
Generically, the two equations in (\ref{HHp})
lead to quadratic equations for $Q$ and $j$ that do not factorise
over the rational functions, thus 
giving rise to rather complicated expressions involving square roots. 
However, in the case of pairwise equal charges the equations do factorise.
The two solutions, which we shall denote by $(j_1,Q_1)$ and $(j_2,Q_2)$,  
are as follows:
%%%%%
\bea
j_1 &=&\fft{r^2+a^2 + 2mr (s_1^2+s_2^2) + 4 m^2 s_1^2 s_2^2}{a}\,,
\qquad
Q_1= -
\fft{(r+ 2 m s_1^2)^2(r+ 2m s_2^2)^2}{a^2}\,,\nn\\
j_2&=& -\fft{r^3-3m r^2 -2m^2 r (s_1^2+s_2^2+2 s_1^2 s_2^2) +
   a^2[m+r + 2m(s_1^2+s_2^2)]}{a(r-m)}\,,\nn\\
Q_2&=&  \fft1{a^2(r-m)^2}\Big\{ -[r^3-3mr^2 -2m^2 
     r(s_1^2+s_2^2 + 2 s_1^2 s_2^2) + 4 m^3 s_1^2 s_2^2]^2 \nn\\
&& \qquad \qquad\qquad+
4 m a^2[r^3(1+s_1^2+s_2^2) +m r^2(s_1^4+s_2^4+ s_1^2+s_2^2 + 6 s_1^2 s_2^2)
\nn\\
&&\qquad\qquad\qquad\qquad\quad 
 -4 m^2 r s_1^2 s_2^2 (1-s_1^2-s_2^2) -
  4m^3 s_1^2 s_2^2 (s_1^2+s_2^2)]\Big\}\,.
\eea
%%%%%

  In the first solution $Q$ is negative, and hence a necessary condition 
for the existence of a solution of the $\theta$ equation is that
%%%%%
\be
 a^2 - Q - j^2 >0\,.
\ee
%%%%
However, for this solution we find
%%%%%
\be
a^2 - Q_1 -j_1^2 =-2(r^2 + 2mr (s_1^2+s_2^2) + 4 m^2 s_1^2 s_2^2)\,,
\ee
%%%%%
which is always negative.  Hence the first solution is unphysical.

\section{Conclusion}

In this paper we have shown that the STU family of rotating charged
black hole solutions of supergravity theory provide a  perfectly manageable
%four  parameter 
family of 
exact metrics   which permit separation
of variables for null and timelike geodesics and are sufficiently explicit
as to allow the calculation of black hole shadows up
to some simple numerical procedures. 
While the most general class of charged 
STU rotating black holes \cite{Chow:2013tia,Chow:2014cca}  also admits a 
separation for both
null and timelike geodesics, we primarily focused on studies of 
the much simpler four-charge parameter black holes  
\cite{Cvetic:1996kv,Chong:2004na}. Furthermore, significant further 
simplifications occur for pairwise-equal charge parameters.
Although we have restricted our attention
to photon orbits and the shadow phenomenon, the utility of these metrics
is not restricted to the behaviour of photon orbits, and we anticipate
further applications in the future.

We should note that classes of rotating black holes in gauged supergravity 
theories also admit a separation for both the null and timelike geodesics, 
which is again a consequence of the separability of the minimally-coupled 
massless scalar equations in these backgrounds, 
c.f., \cite{Birkandan:2015yda}.
Thus the extension of the analysis presented in  this paper to  
supergravity black holes with non-zero cosmological constant 
would provide interesting examples for studying the effects of the cosmological constant in the calculation of the
properties of black hole shadows.

\vskip 0.5cm

\noindent {\large \bf Acknowledgements}
\vskip 0.5cm
We are very grateful to the Mitchell Family Foundation for hospitality at
the Brinsop Court workshop on strings and cosmology, where some of this work
was carried out.
The  research  of  M.C.~is
supported in part by the DOE Grant Award de-sc0013528, the Fay R. and
Eugene  L.  Langberg  Endowed  Chair  and  the  Slovenian  Research  Agency (ARRS).  C.N.P.~is
supported in part by DOE grant DE-FG02-13ER42020.

\appendix

\section{STU Black Holes with Pairwise-Equal Charges}

   In this appendix we collect some results from \cite{Chong:2004na}, where
the STU supergravity rotating 
black holes with pairwise-equal charges were presented.  The bosonic 
Lagrangian describing these black holes is given by\footnote{We have dualised
the field strength $F_1$, relative to the one in \cite{Chong:2004na},
so that both field strengths now carry electric charges in the black 
hole solutions. We also
rescaled the field strengths  so that they each have 
the ``canonical normalisation''
where ${\cal L}\sim \sqrt{-g}(R-F_1^2 - F_2^2)$ in the absence of scalars.}
%%%%%
\bea
{\cal L} &=& \sqrt{-g}\, \Big[R -\ft12 (\del\varphi)^2 -\ft12 e^{2\varphi}\, 
   (\del\chi)^2 - \fft1{(1+\chi^2\, e^{2\varphi})}\, [e^\varphi\,
    F_1^{\mu\nu}\, F_{1\, \mu\nu} -
    \ft12 \chi\, e^{2\varphi}\, \epsilon_{\mu\nu\rho\sigma}\, 
   F_1^{\mu\nu}\, F_1^{\rho\sigma}]\nn\\
&&\qquad\quad
-e^{-\varphi}\, 
    F_2^{\mu\nu}\, F_{2\, \mu\nu} -\ft12
 \chi\, \epsilon_{\mu\nu\rho\sigma}\,
      F_2^{\mu\nu}\, F_2^{\rho\sigma} \Big]\,.\label{pairlag}
\eea
%%%%%
If we define $\tau= \chi + \im\, e^{-\varphi}$ and
%%%%%
\be
F^{\pm}_{1\, \mu\nu}= \fft12 \Big(F_{1\,\mu\nu} \pm \fft{\im}{2}\,
   \epsilon_{\mu\nu\rho\sigma}\, F_1^{\rho\sigma}\Big)\,,\qquad
F^{\pm}_{2\, \mu\nu}= \fft12 \Big(F_{2\,\mu\nu} \pm \fft{\im}{2}\,
   \epsilon_{\mu\nu\rho\sigma}\, F_2^{\rho\sigma}\Big)\,, 
\ee
%%%%%%
then the Lagrangian (\ref{pairlag}) can be written as
%%%%%
\be
{\cal L}= \sqrt{-g}\, \Big\{ R - 
 \fft{|\del\tau|^2}{2 \tau_2^2} 
  -2\, \Im \Big[\tau\, F^{+\,\mu\nu}_2\, F^{+}_{2\, \mu\nu} +
  \Big(-\fft1\tau\Big)\, F^{+\,\mu\nu}_1\, F^{+}_{1\, \mu\nu}\Big]\Big\}\,,
\ee 
%%%%%
where $\tau_2= \Im(\tau)= e^{-\varphi}$.  The Bianchi identities and
equations of motion for the field strengths can be written as
%%%%%%
\be
  \nabla_\mu \Im(F_1^{+\, \mu\nu}) =  0\,,\qquad
\nabla_\mu \Im(F_2^{+\, \mu\nu}) =  0
\ee
%%%%%
and 
\be
  \nabla_\mu \Im(G_1^{+\, \mu\nu}) =  0\,,\qquad
\nabla_\mu \Im(G_2^{+\, \mu\nu}) =  0
\ee
%%%%%
respectively, where
%%%%%
\be
G^+_{1\, \mu\nu}= \Big(-\fft1\tau\Big)\, F^+_{1\, \mu\nu}\,,\qquad
G^+_{2\, \mu\nu}= \tau\, F^+_{2\, \mu\nu}\,.
\ee
%%%%%
The Bianchi identities and equations of motion transform covariantly
under the global $SL(2,R)$ symmetry
%%%%%
\be
\tau\rightarrow \fft{a\tau+b}{c\tau+d}\,,\qquad
\begin{pmatrix} F^+_1\cr G^+_1\end{pmatrix} \rightarrow
 \begin{pmatrix}a & -b\cr -c & d\end{pmatrix} \, 
  \begin{pmatrix} F^+_1\cr G^+_1\end{pmatrix} \,,\qquad
\begin{pmatrix} F^+_2\cr G^+_2\end{pmatrix} \rightarrow
 \begin{pmatrix}d & c\cr b & a\end{pmatrix} \,
  \begin{pmatrix} F^+_2\cr G^+_2\end{pmatrix} \,,
\ee
%%%%%
where $ad -bc = 1$.  Note that this is a symmetry only at the level
of the equations of motion, and not of the Lagrangian. There is, in
addition, a discrete $Z_2$ symmetry of both the equations of motion and the
Lagrangian, under 
%%%%%
\be
\tau\rightarrow -\fft1{\tau}\,,\qquad A_1\rightarrow A_2\,,\qquad
  A_2\rightarrow A_1\,.
\ee
%%%%%

The metric, dilationic scalar, axion and gauge potentials are given by
%%%%%
\bea
ds^2 &=& W\, \Big(\fft{dr^2}{\Sigma} + d\theta^2\Big)
  -\fft{\Sigma}{W}\, [dt- a \sin^2\theta\, d\phi]^2 +
   \fft{\sin^2\theta}{W}\, [a dt -(r_1 \, r_2 + a^2) d\phi]^2\,,\nn\\
e^{\varphi} &=& \fft{r_1^2 + a^2 \cos^2\theta}{r_1\, r_2 + a^2\cos^2\theta}
\,,\qquad \chi = 
 \fft{2 m a (s_2^2 - s_1^2)\, \cos\theta}{r_1^2 + a^2\cos^2\theta}\,,\nn\\
A_1 &=& \fft{\sqrt2\, m s_1 c_1\, r_2\, [dt - a\sin^2\theta\, d\phi]}{W}\,,
\qquad
A_2= \fft{\sqrt2\, m s_2 c_2\, r_1\, [dt - a\sin^2\theta\, d\phi]}{W}\,,
\label{pairwisesol}
\eea
%%%%%
where
%%%%%
\bea
\Sigma &=& r^2 - 2mr + a^2 \,,\qquad W = r_1\, r_2 + a^2\cos^2\theta\,,\nn\\
r_1 &=& r + 2 m s_1^2\,,\qquad r_2= r + 2 m s_2^2\,.
\eea
%%%%%
As usual, $s_i=\sinh\delta_i$, $c_i=\cosh\delta_i$, and the two constants
$\delta_1$ and $\delta_2$ parameterise the electric charges carried by the
two field strengths $F_{i\, \mu\nu}$. 
The mass, angular momentum and electric charges 
(with the field strengths normalised as in (\ref{pairlag})) are
given by
%%%%%
\be
M=\fft{m}{2}\, (\cosh2\delta_1+\cosh2\delta_2)\,,\qquad
J= \fft{m a}{2}\, (\cosh2\delta_1+\cosh2\delta_2)\,,\qquad
Q_i= \fft{m}{\sqrt2}\, \sinh2\delta_i\,.
\ee
%%%%%

\end{document}